\RequirePackage{fix-cm}

\documentclass[smallextended]{svjour3}       % onecolumn (second format)
\usepackage{amsmath}
\usepackage{amssymb}

\smartqed  % flush right qed marks, e.g. at end of proof
\usepackage{graphicx}
\begin{document}

\title{Discrete phase space and continuous time relativistic quantum mechanics I: Planck oscillators and closed string-like circular orbits}

\author{Anadijiban Das         \and
        Rupak Chatterjee     
}

\institute{Anadijiban Das \at
              Department of Mathematics, Simon Fraser University, Burnaby, British Columbia, V5A 1S6, Canada \\
              \email{Das@sfu.ca}           %  \\
           \and
           Rupak Chatterjee \at
              Center for Quantum Science and Engineering\\
Department of Physics, Stevens Institute of Technology, Castle Point on the Hudson, Hoboken, NJ 07030, USA \\ \email{Rupak.Chatterjee@Stevens.edu}   
}

\date{Received: date / Accepted: date}

\maketitle

\begin{abstract}
The discrete phase space continuous time representation of relativistic quantum mechanics involving a characteristic length $l$ is investigated. Fundamental physical constants such as $\hbar$, $c$, and $l$ are retained for most sections of the paper. The energy eigenvalue problem for the Planck oscillator is solved exactly in this framework. Discrete concircular orbits of constant energy are shown to be circles $S^{1}_{n}$ of radii $2E_n = \sqrt{2n+1}$ within the discrete $(1+1)$-dimensional phase plane. Moreover, the time evolution of these orbits sweep out world-sheet like geometrical entities $S^{1}_{n} \times \mathbb{R} \subset \mathbb{R}^2 $ and therefore appear as closed string-like geometrical configurations. The  physical interpretation for these discrete orbits in phase space as degenerate, string-like phase cells is shown in a mathematically rigorous way. The existence of these closed concircular orbits in the arena of discrete phase space quantum mechanics, known for the non-singular nature of lower order expansion $S^{\#}$ matrix terms, was known to exist but has not been fully explored until now. Finally, the discrete partial difference-differential Klein-Gordon equation is shown to be invariant under the continuous inhomogeneous orthogonal group $\mathcal{I}\left[O(3,1)\right]$. 


\end{abstract}

\section{Introduction}
There have been many attempts in the past to quantize space-time in order to remove the divergences found in quantum field theory. In 1960, a divergence free quantum field theory was presented in the arena of lattice space-time \cite{DasI}. Therein, the mathematics of partial difference field equations was used but unfortunately relativistic covariance was not yet possible.

Traditionally, quantum mechanics has been formulated in a space-time continuum \cite{Greiner}. It is well known that quantum mechanics may be represented in terms of phase space variables within a phase space continuum \cite{Weyl,WignerI} often involving a characteristic length scale $l > 0$. The analog of the “wave function” in this formulation is the  Wigner function being a probability distribution function in phase space. Furthemore, a charactersitic length scale has also been utilized by  Heisenberg in his pioneering work on matrix mechanics \cite{Heisenberg}. After many decades, partial difference operators involving a characteristic length scale $l > 0$ were utilized in an exact quantum mechanical wave mechanics formulation that maintained exact relativistic covariance \cite{DasII,DasIII}. Following this was the formulation of the discrete space and continuous time $S^{\#}$-matrix theory of interacting fields \cite{DasIV,DasV,DasVI} whereby the special relativistic $S^{\#}$-matrices were found to be non-singular for many lower order expansion terms.

In the year 2015, three-dimensional Planck oscillators \cite{DasVII} were introduced in a $(2+2+2)$-dimensional phase space continuum (i.e. $\{(p_x,q_x),(p_y,q_y),\cdots \}$) formulation of quantum field theory. In this paper,  we concentrate on partial difference equations in a discrete phase space being a subset of the continuous phase space. In the wave mechanical version of the Planck oscillator, discrete orbits $ S^{1}_{n^1} \times S^{1}_{n^3} \times S^{1}_{n^3} $ emerge inside the $(2+2+2)$-dimensional phase space where $ S^{1}_{n}$ indicates a one-dimensional circle of radius $\sqrt{2n+1}$ with its center at the origin of a two-dimensional phase plane. We study these concircular orbits and the direct products of these circles in sections 3,4, and 5. Relativistic second quantized waves fields are shown to have quanta inside the $(2+2+2)$ discrete phase space characterized by these direct product orbits. Moreover, use of circular hyper-cylinders  $ S^{1}_{n^1} \times S^{1}_{n^3} \times S^{1}_{n^3} \times \mathbb{R}$ of discrete phase space and continuous time are explored in sections 5, 6, and 7. 
 
In a previous work \cite{DasVI}, quantum electrodynamics involving a $S^{\#}$-matrix was derived in the arena of $[(2+2+2)+1]$-dimensional discrete phase space and continuous time. Many of the lower order expansion terms of the $S^{\#}$-matrix turned out to be non-singular as opposed to the usual case of divergences seen in the traditional continuum methods of quantum field theory. Furthermore, a physically verifiable non-singular Coulomb potential was derived from a discrete quantum electrodynamics model \cite{DasVIII}. 

The circle $ S^{1}_{n}$ inside the $(1+1)$ dimensional discrete phase space has been shown to resemble a string-like object \cite{DasIX,Zwiebach,Polchinski} while the direct product $ S^{1}_{n^1} \times S^{1}_{n^3} \times S^{1}_{n^3}$ inside the $(2+2+2)$-dimensional discrete phase space resembles a hyper-spherical torus  \cite{Zwiebach,Polchinski}. This topic will be treated in greater detail in a follow up (part II) to the current work.

The discrete phase space and continuous time model can incorporate the second quantized form of both free and interacting relativistic fields. Similar formulations can also include Einstein's linearized gravitational theory \cite{Einstein}. 

Finally, in section 7, we consider world sheet like geometrical figures given by the Cartesian direct product $ S^{1}_{n^1} \times S^{1}_{n^3} \times S^{1}_{n^3} \times \mathbb{R}$ in the arena of $[(2+2+2)+1]$-dimensional discrete phase space and continuous time. Specifically, we discuss the relativistic Klein-Gordon wave equation in such a space time. The relativistic invariance  of the partial difference-differential discretization of the Klein-Gordon equation is proved rigorously. The Klein-Gordon wave function will be denoted as $\phi(n^1, n^2, n^3, t)=:\phi(\mathbf{n},t)$. The second quantized version of this wave function will be denoted as $\Phi(\mathbf{n},t)$. We assume that the massive particle quanta of $\Phi(\mathbf{n},t)$ can cohabit with the quantized Planck oscillators inside the geometrical configuration of $ S^{1}_{n^1} \times S^{1}_{n^3} \times S^{1}_{n^3} \times \mathbb{R}$ temporarily or forever.

\section{Classical and Quantum Planck Oscillators in a (1+1)-dimensional Phase Plane Continuum $\mathbb{R}^2$}

We will explicitly use physical units $\hbar = \left[ \dfrac{ML^2}{T} \right], c=\left[ \dfrac{L}{T} \right]$, and $l={L}$ here where the characteristic length $l >0 $ is to be set by future experimental results. However, we expect $l$ to be close to the Planck length $\sqrt{\dfrac{\hbar G}{c^3}}$. The classical Planck oscillator \cite{DasVII} in the $(1+1)$-dimensional phase plane continuum is provided by the physically dimensionless Hamiltonian
\begin{equation}
H(q,p):= \left( \dfrac{1}{2} \right) \left[ \left( \dfrac{l}{\hbar} \right)^2 p^ 2 + \left( \dfrac{1}{l} \right)^2 q^2 \right] = \left[ \dfrac{E}{\hbar \nu} \right]
\end{equation} 
where $E$ and $\nu$ stand for the total energy and circular frequency of the classical Planck oscillator respectively.

The graph of the classical motion in the phase plane $\mathbb{R}^2$ is depicted by the ellipse in figure 1. Note that in the limit $l \rightarrow 0^{+}$, the ellipse collapses into the $p$-axis.
\begin{figure}
\begin{center}
\includegraphics[scale=0.35]{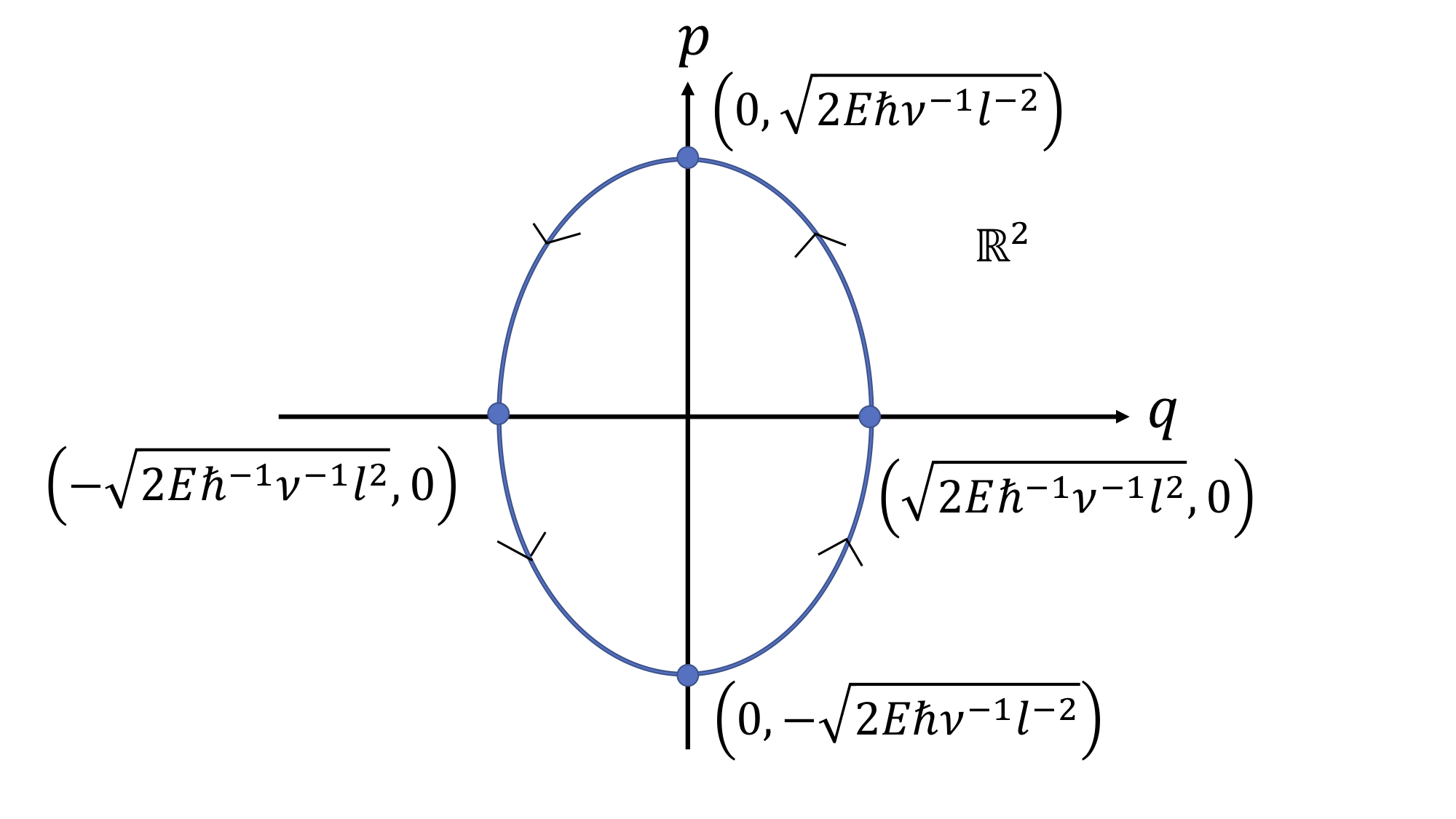}
\end{center}
\caption{Ellipse Representing the Classical Trajectory of the Planck Oscillator}
\end{figure}
The elliptical world-tube \cite{Synge} in figure 2 represents the {Planck oscillator in the $[(1+1)+1]$-dimensional state space ($(p,q,t)$) \cite{Lanczos}.
\begin{figure}
\begin{center}
\includegraphics[scale=0.35]{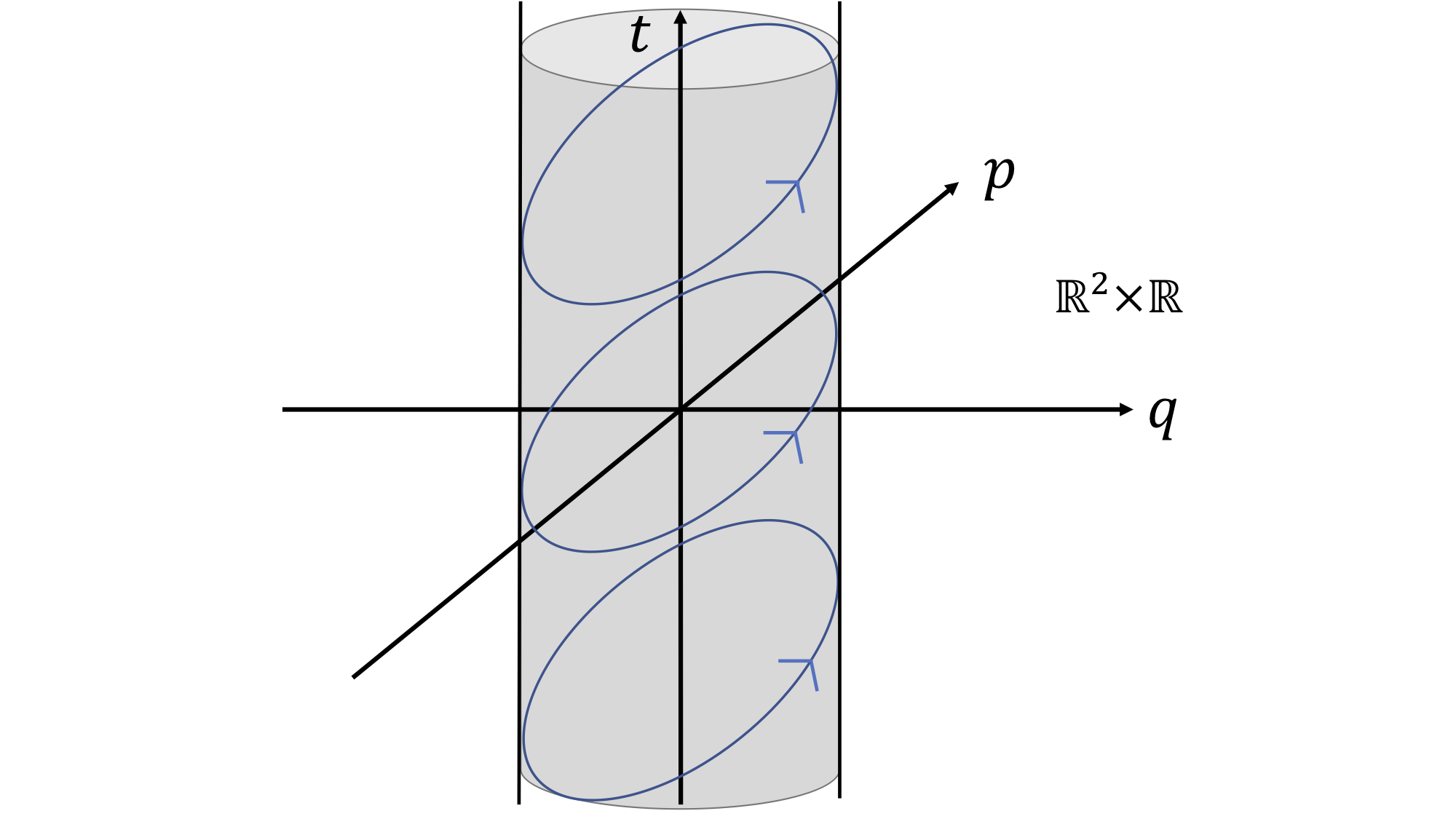}
\end{center}
\caption{Elliptical Vertical Cylinder inside the $[(1+1)+1]$-dimensional State Space of the Planck Oscillator}
\end{figure}
In figure 1, the elliptical trajectory resembles a closed string-like configuration. On the other hand, the elliptical vertical cylinder in figure 2 resembles an open world sheet like object \cite{Zwiebach,Polchinski} inside the $[(1+1)+1]$-dimensional phase plane plus time continuum $\mathbb{R}^2 \times \mathbb{R}$. 

The quantum mechanical solution of the Planck oscillator is straightforward and is as follows. The state vector $\overrightarrow{\mathbf{\Psi}}$ and linear operators $\mathbf{Q}$ and $\mathbf{P}$ in a separable Hilbert space \cite{Lorch} renders the classical Hamiltonian (1) into the quantum eigenvalue problem 
 \begin{equation}
H(\mathbf{Q},\mathbf{P}) \overrightarrow{\mathbf{\Psi}} = \left( \dfrac{1}{2} \right) \left[ \left( \dfrac{l}{\hbar} \right)^2 \mathbf{P}^ 2 + \left( \dfrac{1}{l} \right)^2 \mathbf{Q}^2 \right] \overrightarrow{\mathbf{\Psi}} = \left[ \dfrac{E}{\hbar \nu} \right] \overrightarrow{\mathbf{\Psi}}.
\end{equation}
The well-known Schr\"{o}dinger representation \cite{Schiff} is furnished by
\begin{equation}
\overrightarrow{\mathbf{\Psi}} := \psi(q), \,\,\, \mathbf{Q}\overrightarrow{\mathbf{\Psi}}=q \psi(q), \,\,\,  \mathbf{P}  \overrightarrow{\mathbf{\Psi}}:= -i\hbar \dfrac{d}{dq} \psi(q).
\end{equation}
Using the above, the Schr\"{o}dinger wave equation follows as 
\begin{equation}
\dfrac{d^2\psi(q)}{dq^2} +\left[\left( \dfrac{2E}{\hbar \nu} \right) - \left( \dfrac{q}{l} \right)^2 \right] \psi(q)=0.
\end{equation}
with a solution set given by \cite{Schiff}
\begin{equation}
\begin{array}{c}
E_N = \left( N +\dfrac{1}{2} \right) \hbar \nu, \,\,\, N \in \{0,1,2,...\}; \\
\\
\psi_N (q/l) =\dfrac{\exp[(-q/l)^2/2]}{\pi^{1/4}2^{N/2}\sqrt{N! l}} H_N(q/l), \\
\\
\overrightarrow{\mathbf{\Psi}}_N \cdot \overrightarrow{\mathbf{\Psi}}_M =\int_{\mathbb{R}} \bar{\psi}_N (q/l) {\psi}_M (q/l) dq = \delta_{NM}.
\end{array}
\end{equation}
where $H_N(q/l)$ are Hermite polynomials \cite{Schiff,Gradshteyn}. Note that the last expression in (5) is physically dimensionless.

\section{Planck Oscillators in a new coordinate system of the (1+1)-dimensional Phase Plane Continuum $\mathbb{R}^2$}

Let us introduce a new coordinate system of physically dimensionless quantities for the $(1+1)$-dimensional phase plane continuum in $\mathbb{R}^2$ as follows (hatted quantities refer to transformed versions of the corresponding unhatted quantities)
\begin{equation}
\begin{array}{c}
y= \left( \dfrac{1}{2} \right)  \left( \dfrac{q}{l} \right)^2 > 0,\\
\\
\left( \dfrac{q}{l} \right) = +\sqrt{2y} > 0.
\end{array}
\end{equation}
This leads to the following transformation of (3), 
\begin{equation}
\begin{array}{c}
\overrightarrow{\mathbf{\Psi}} = \hat{\psi}(y), \\
\\
\mathbf{Q} \overrightarrow{\mathbf{\Psi}}=(\sqrt{2y} l ) \hat{\psi}(y), \\
\\
\mathbf{P} \overrightarrow{\mathbf{\Psi}}:=  -i\left( \dfrac{\hbar}{l} \right) \sqrt{2y} \dfrac{d}{dy}\hat{\psi}(y)
\\ 
\\
= -i\left( \dfrac{\hbar}{l} \right) \sqrt{\dfrac{y}{2}} \left\{  \displaystyle{\lim_{a \to 0_+}} \left[\dfrac{\hat{\psi}(y+a)-\hat{\psi}(y)}{a} \right] + \displaystyle{\lim_{a \to 0_-}} \left[\dfrac{\hat{\psi}(y+a)-\hat{\psi}(y)}{a} \right] \right\}.
\end{array}
\end{equation}
The Schr\"{o}dinger wave equation is furnished by
\begin{equation}
\sqrt{2y} \dfrac{d}{dy} \left[ \sqrt{2y} \dfrac{d}{dy}\hat{\psi}(y) \right] + \left[ \left( \dfrac{2E}{\hbar \nu}\right) -2y \right] \hat{\psi}(y) =0,
\end{equation}
with eigenvalue solutions of 
\begin{equation}
\begin{array}{c}
E_N = \left( N +\dfrac{1}{2} \right) \hbar \nu ,\\
\\
\hat{\psi}_N (y) = \dfrac{e^{-y}H_N(\sqrt{2y})}{\pi^{1/4}2^{N/2}\sqrt{N!}}.
\end{array}
\end{equation}
The real valued wave function $\hat{\psi}_N (y)$ is physically dimensionless as desired.

\section{Quantum Planck Oscillators in a (1+1)-dimensional Discrete Phase Plane}

Let us begin by defining various finite difference operations \cite{DasIV,DasVII,DasIX} on a real or complex-valued wave function $\phi(n)$ as 
\begin{equation}
\begin{array}{c}
\Delta \phi(n) := \phi(n+1)-\phi(n) ,\\
\\
\Delta^{'} \phi(n) := \phi(n)-\phi(n-1) ,\\
\\
\Delta^{o} \phi(n) := \left(\dfrac{1}{\sqrt{2}} \right)\left[ \sqrt{n+1} \phi(n+1) +\sqrt{n} \phi(n-1) \right],\\
\\
\Delta^{\#} \phi(n) := \left(\dfrac{1}{\sqrt{2}} \right)\left[ \sqrt{n+1} \phi(n+1) -\sqrt{n} \phi(n-1) \right],\\
\\
n \in \{0,1,2,...\}.
\end{array}
\end{equation}
There exists many more variations of difference operators (see \cite{Jordan}). $\Delta^{\#}$ has beed used several times before \cite{DasIV,DasVII,DasVIII,DasIX} whereas $\Delta^{o}$ may be found in \cite{DasVII,DasIX}. Mathematicians generally use finite difference operators for numerical analysis. They are generally not concerned with the physical dimensions of various mathematical entities. Yet, for our purposes, we must be careful at being able to return to physical dimensions for experimental verification once we go dimensionless. 

Consider a scale $l >0$ without concern for the physical dimensionality such that the mathematical analysis reads as follows,
\begin{equation}
\begin{array}{c}
y_n := nl, \,\,\, y_{n+1}-y_{n}=l, \\
\\
\chi(nl)  \equiv \chi(y_n), \\
\\
\chi(\dfrac{y_n}{l}) =: \phi(n), \\
\\
\Delta_l \left[ \chi(\dfrac{y_n}{l}) \right] =: \Delta \phi(n), \\
\\
\Delta^{'}_l \left[ \chi(\dfrac{y_n}{l}) \right] =: \Delta^{'} \phi(n), \\
\\
\Delta^{o}_l \left[ \chi(\dfrac{y_n}{l}) \right] =: \left( \dfrac{l}{2} \right)^{1/2} 
\left[ \sqrt{(n+1)l} \phi(n+1) +\sqrt{nl} \phi(n-1) \right] = l \Delta^{o} \phi(n), \\
\\
(-i\hbar) \Delta^{\#}_l \left[ \chi(\dfrac{y_n}{l}) \right] =: (-i\hbar) \left(\dfrac{1}{\sqrt{2}l^{3/2}} \right)\left[ \sqrt{n+1} \phi(n+1) -\sqrt{n} \phi(n-1) \right] \\
\\
=(-i\hbar) \left( \dfrac{1}{l} \right)\Delta^{\#} \phi(n). 
\end{array}
\end{equation}

Returning to the physically admissible first quanitized Planck oscillator model, let the state vector $\overrightarrow{\mathbf{\Psi}}$ and linear operators $\mathbf{Q}$ and $\mathbf{P}$ be defined in the following manner:
\begin{equation}
\begin{array}{c}
\overrightarrow{\mathbf{\Psi}} := \chi(\dfrac{y_n}{l}) := \phi(n), \\
\\
\mathbf{Q} \overrightarrow{\mathbf{\Psi}} := \Delta^{o}_l \chi(\dfrac{y_n}{l}) = l \Delta^{o} \phi(n),\\
\\
\mathbf{P} \overrightarrow{\mathbf{\Psi}}:=  -i\hbar \Delta^{\#}_l \chi(\dfrac{y_n}{l}) = \left( \dfrac{-i\hbar}{l} \right)\Delta^{\#} \phi(n).
\end{array}
\end{equation}
The first equation of (12) is physically dimensionless whereas the second and third equations are of physical dimensions of length and momentum respectively. From these, we can directly compute the fundamental commutation relation,
\begin{equation}
[\mathbf{P}, \mathbf{Q}] \overrightarrow{\mathbf{\Psi}} := (\mathbf{P} \mathbf{Q} - \mathbf{Q} \mathbf{P}) \overrightarrow{\mathbf{\Psi}} = (-i \hbar) (\Delta^{\#}\Delta^{o}-\Delta^{o}\Delta^{\#})\phi(n)=-i\hbar \phi(n)= -i\hbar \overrightarrow{\mathbf{\Psi}}.
\end{equation}
The quantum Hamiltonian equation (2) reduces to 
\begin{equation}
\begin{array}{c}
\left( \dfrac{1}{2} \right) \left[ \left( \dfrac{l}{\hbar} \right)^2 \mathbf{P}^ 2 + \left( \dfrac{1}{l} \right)^2 \mathbf{Q}^2 \right] \overrightarrow{\mathbf{\Psi}} = \left[ \dfrac{E}{\hbar \nu} \right] \overrightarrow{\mathbf{\Psi}}, \\ 
or \\
\left[-(\Delta^{\#})^2 +(\Delta^{o})^2\right] \phi(n) = \left( \dfrac{2E}{\hbar \nu} \right)\phi(n).
\end{array}
\end{equation}
The corresponding eigenvalue problem is exactly solvable with the following solution
\begin{equation}
\begin{array}{c}
E_N =\left(N+\dfrac{1}{2}\right)\hbar \nu , \,\,\, E_0=\dfrac{\hbar \nu}{2}, \\
\\
N \in\{0,1,2,...\},\\ 
\\
\overrightarrow{\mathbf{\Psi}}_N=\phi_N(n)=\delta_{Nn}, \\
\\
\overrightarrow{\mathbf{\Psi}}_N^{\dagger} \cdot \overrightarrow{\mathbf{\Psi}}_M := \displaystyle{\sum_{n=0}^{\infty}} \delta_{Nn} \delta_{Mn}=\delta_{NM}, \\
\\
||\overrightarrow{\mathbf{\Psi}}_N||^2 =1.
\end{array}
\end{equation}
Equations (13) and (15) prove that the present finite difference representations of quantum mechanics are exact (with the simplest possible eigenfunction ever of $\phi_N(n)=\delta_{Nn}$). By (14) and (15), the discretization of the phase plane via the relation  $\left( \dfrac{1}{2} \right) \left[ \left( \dfrac{l}{\hbar} \right)^2 p^ 2 + \left( \dfrac{1}{l} \right)^2 q^2 \right] = \left(n+\dfrac{1}{2}\right), \,\,\, n\in\{0,1,2,...\}$ is achieved. This relation represents denumerably infinite numbers of confocal ellipses in the background of the continuous $q-p$ phase plane \cite{DasVII}. Therefore, the particle’s accessible phase space is discrete in this case ('phase cell') in such a way that the relation between $q$ and $p$ is integer valued.

Finally, from (11) and (12), we have
\begin{equation}
\begin{array}{c}
\mathbf{Q} \overrightarrow{\mathbf{\Psi}} := \sqrt{2y_n l} \,\,\chi\left(\dfrac{y_n}{l}\right) + \left[ \Delta \left( \sqrt{\dfrac{y_n l}{2}}\,\chi\left(\dfrac{y_n}{l}\right) \right) -\sqrt{\dfrac{y_n l}{2}} \Delta^{'}\left(\chi\left(\dfrac{y_n}{l}\right)\right) \right], \\
\\
\mathbf{P} \overrightarrow{\mathbf{\Psi}}:= -i\hbar \left(\dfrac{1}{l} \right) \left[ \Delta \left( \sqrt{\dfrac{y_n l}{2}}\,\chi\left(\dfrac{y_n}{l}\right) \right) +\sqrt{\dfrac{y_n l}{2}} \Delta^{'}\left(\chi\left(\dfrac{y_n}{l}\right)\right) \right].
\end{array}
\end{equation}

With respect to the first quantized Planck oscillator equations in the unusual coordinate system of section 3, a mathematician conversed in standard numerical analysis 
\cite{Jordan,Pettofrezzo} will the finite difference equations involving operators $\Delta$ and $\Delta^{'}$ only to obtain approximate behaviors of the unknown functions $\phi(n) \equiv \chi \left(\dfrac{y_n}{l} \right)$. Yet, our representation of quantum mechanics using the operators  $\Delta^{\#}$ and $\Delta^{o}$ provide the exact quantum conditions (13)-(15). Note also the striking similarity between the differential relations found in (7) and the difference relations found above in (16).

\section{String-like Phase Plane Orbits and Fundamental Units}

The fundamental physical units in this section will be chosen to be $\hbar = 1 \left[ \dfrac{ML^2}{T} \right], c = 1  \left[ \dfrac{L}{T} \right],l = 1 [L]$. In these units, the discrete phase plane 'phase cell' is denoted by $S^{1}_n$. The mathematical equations of the previous sections reduce to much simpler forms if expressed in these fundamental units. The one-dimensional circular orbits $S^{1}_N$ with discrete energy $2E_N = (2N+1)$ are expressed as, following (15), 
\begin{equation}
\begin{array}{c}
S^{1}_N := \left\{ (q,p) \in \mathbb{R}^2 : q^2+p^2=2N+1 \in \mathbb{R} \right\}, \\
\\
S^{1}_n \equiv S^{1}_N,\\
\\
N \in \{0,1,2,\cdots \}.
\end{array}
\end{equation}
These circular orbits are depicted in figure 3.
\begin{figure}
\begin{center}
\includegraphics[scale=0.35]{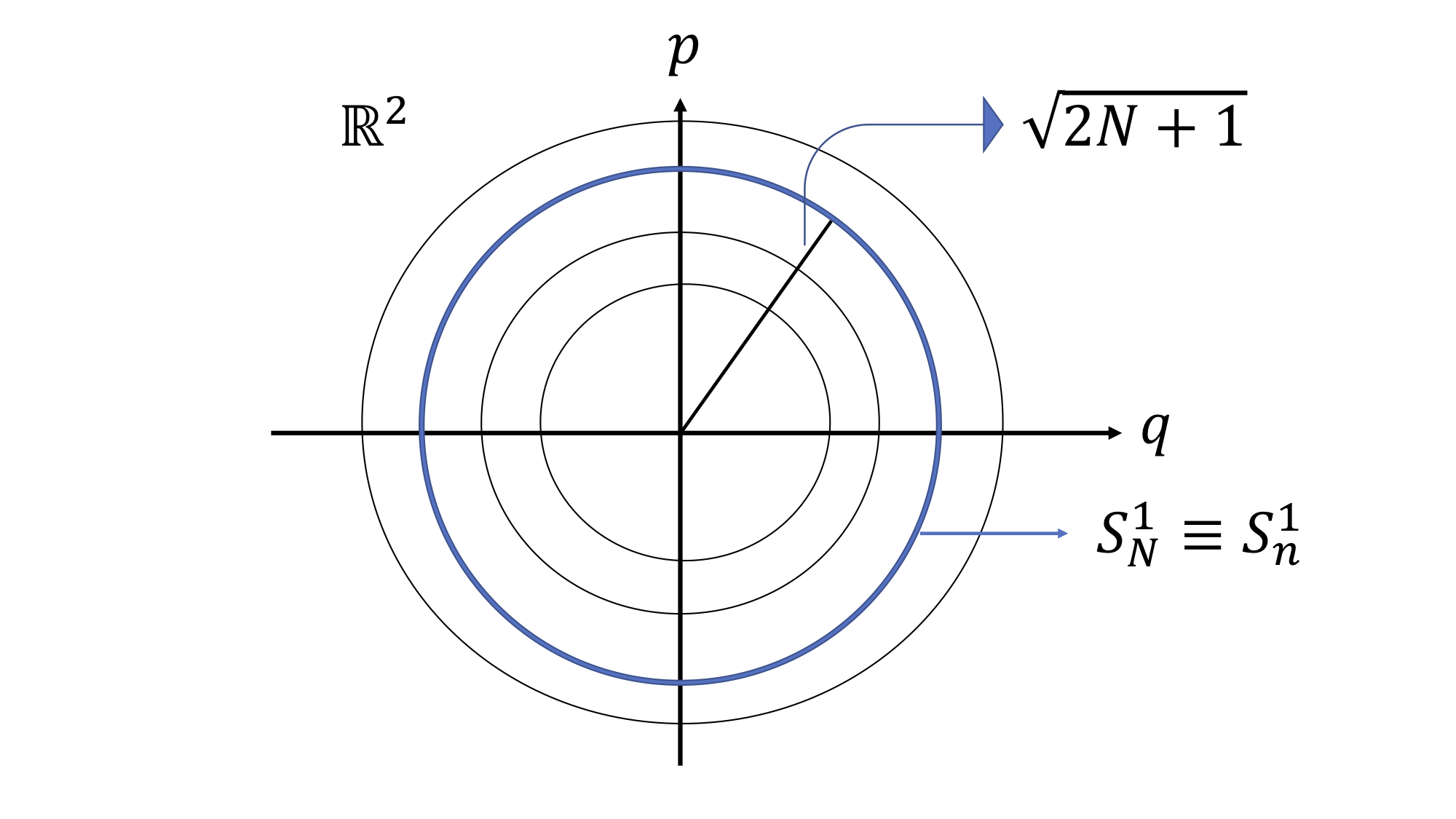}
\end{center}
\caption{The String-like Closed Circular Orbit $S^{1}_N$ for the First Quantized Planck Oscillator}
\end{figure}
The corresponding $(1+1)$-dimensional world-sheet like circular cylinder \cite{Zwiebach} in the $[(1+1)+1]$-dimensional discrete phase plane and continuous time is exhibited in figure 4.
\begin{figure}
\begin{center}
\includegraphics[scale=0.35]{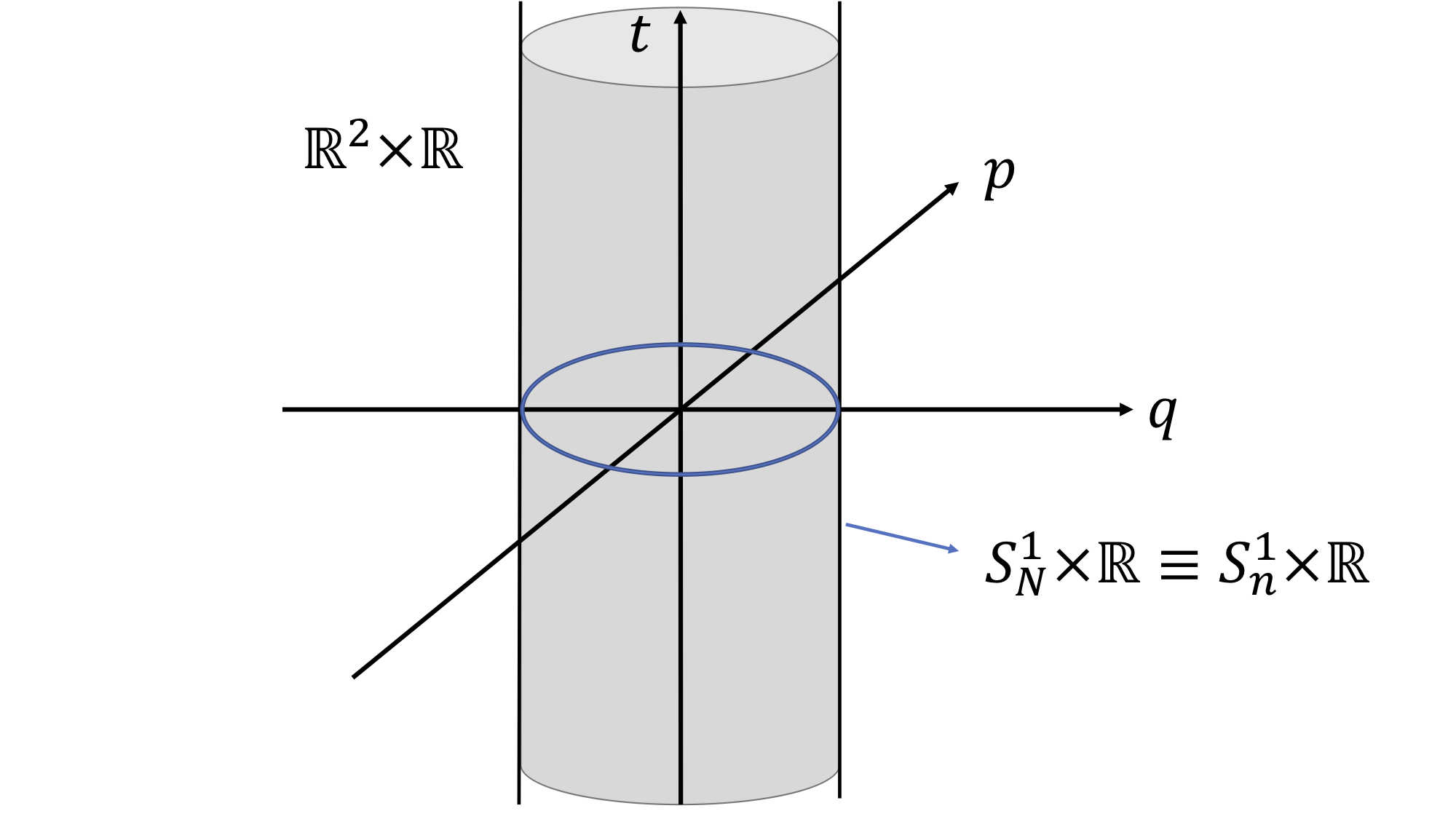}
\end{center}
\caption{The World-Sheet like Circular Cylinder $S^{1}_N \times \mathbb{R}$ embedded inside the $[(1+1)+1]$-dimensional State Space (discrete phase plane and continuous time)}
\end{figure}

These two figures may be compared and contrasted to those of standard string theory \cite{Zwiebach,Polchinski}. The discrete one-dimensional orbit $S^{1}_N$ of figure 3 closely resembles closed strings. There exists a denumerable infinite number of concircular orbits of Planck oscillators in the discrete phase plane. Furthermore, the hollow circular cylinders in state space $S^{1}_N \times \mathbb{R}$ resemble the two-dimensional world sheets of usual string theory. Clearly, a denumerable infinite number of vertical co-axial world-sheet like objects $S^{1}_N \times \mathbb{R}$ accommodate the first quantized Planck oscillators for all time.

\section{Partial Difference and Partial Difference-Differential versions of the Relativistic Klein-Gordon Equation}

The same fundamental physical units of the previous section are used here. Greek indices will take values from $\{1,2,3,4\}$ whereas the Roman indices take values from  $\{1,2,3\}$. Hatted quantities
refer to transformed versions of the corresponding unhatted quantities. The flat space-time metric with signature $+2$ is denoted as $\eta_{\mu \nu}:=[1,1,1,-1]$. An element of the relativistic discrete phase space will be denoted as $(n^1, n^2, n^3, n^4) \in \mathbb{N}^4$ where $n^{\mu}$ is a non-negative integer $\mathbb{N} =\{0,1,2, ...\}$. Moreover, an element of discrete phase space and continuous time will be denoted as $(n^1, n^2, n^3, t) =: (\mathbf{n},t)$. Here $n^j$ is a non-negative integer and $t$ is a real number. The complex or real valued wave function of the relativistic Klein-Gordon equation is denoted as either $\phi(n^1, n^2, n^3, n^4)$ or  $\phi(n^1, n^2, n^3, t) =: \phi(\mathbf{n},t)$.

The various partial difference and differential operations from equation (10) are defined on the relativistic wave function as follows \cite{DasIV,DasVII,DasVIII,DasIX}:
\begin{equation}
\begin{array}{c}
\Delta_{\mu} \phi(n^1, n^2, n^3, n^4) := \phi(\cdots, n^{\mu}+1, \cdots)-\phi(\cdots, n^{\mu}, \cdots), \\
\\
\Delta^{'}_{\mu} \phi(n^1, n^2, n^3, n^4) := \phi(\cdots, n^{\mu}, \cdots)-\phi(\cdots, n^{\mu}-1, \cdots) ), \\
\\
\Delta^{o}_{\mu} \phi(n^1, n^2, n^3, n^4):= \\ 
\left(\dfrac{1}{\sqrt{2}} \right)\left[ \sqrt{n^{\mu}+1} \phi(\cdots, n^{\mu}+1, \cdots) +\sqrt{n^{\mu}} \phi(\cdots, n^{\mu}-1, \cdots) \right],\\
\\
\Delta^{\#}_{\mu} \phi(n^1, n^2, n^3, n^4):= \\
\left(\dfrac{1}{\sqrt{2}} \right)\left[ \sqrt{n^{\mu}+1} \phi(\cdots, n^{\mu}+1, \cdots)  - \sqrt{n^{\mu}} \phi(\cdots, n^{\mu}-1, \cdots) \right].\\\
\end{array}
\end{equation}

Similarly, we define
\begin{equation}
\begin{array}{c}
\Delta_{j} \phi(n^1, n^2, n^3 ;t) := \phi(\cdots, n^{j}+1, \cdots ;t)-\phi(\cdots, n^{j}, \cdots ;y), \\
\\
\Delta^{'}_{j} \phi(n^1, n^2, n^3 ;t) := \phi(\cdots, n^{j}, \cdots ;t)-\phi(\cdots, n^{j}-1, \cdots ;t) ), \\
\\
\Delta^{o}_{j} \phi(n^1, n^2, n^3 ;t):= \\
\left(\dfrac{1}{\sqrt{2}} \right)\left[ \sqrt{n^{j}+1} \phi(\cdots, n^{j}+1, \cdots ;t) +\sqrt{n^{j}} \phi(\cdots, n^{j}-1, \cdots ;t) \right],\\
\\
\Delta^{\#}_{j} \phi(n^1, n^2, n^3 ;t):= \\
\left(\dfrac{1}{\sqrt{2}} \right)\left[ \sqrt{n^{j}+1} \phi(\cdots, n^{j}+1, \cdots ;t)  - \sqrt{n^{j}} \phi(\cdots, n^{j}-1, \cdots ;t) \right],\\
\\
\partial_t \phi(n^1, n^2, n^3 ;t) \equiv \partial_t \phi(\mathbf{n},t) := \dfrac{\partial}{\partial t} \left[ \phi(n^1, n^2, n^3 ;t) \right].
\end{array}
\end{equation}

\section{The Relativistic Klein-Gordon Equation as a Finite Difference or Finite Difference-Differential Equation}

We will know define the quantum or wave version of the relativistic Klein-Gordon equation in the covariant discrete phase space of $(4+4)$-dimension as follows,
\begin{equation}
\begin{array}{c}
\overrightarrow{\mathbf{\Psi}}  := \phi(n^1, n^2, n^3 ;t), \\
\\
\mathbf{Q}^{\mu} \overrightarrow{\mathbf{\Psi}} :=  \eta^{\mu \nu} \Delta^{o}_{\nu} \phi(n^1, n^2, n^3, n^4),\\
\\
\mathbf{P}_{\nu} \overrightarrow{\mathbf{\Psi}} := -i \Delta^{\#}_{\nu} \phi(n^1, n^2, n^3, n^4), \\
\\
\left[ \left( \eta^{\mu \nu} \mathbf{P}_{\mu} \mathbf{P}_{\nu} \right) +m^2 \mathbf{I} \right] \overrightarrow{\mathbf{\Psi}} =\overrightarrow{\mathbf{0}}, \\
or \\
\left[ \left( \eta^{\mu \nu} \Delta^{\#}_{\mu} \Delta^{\#}_{\nu} \right) - m^2 \right] \phi(n^1, n^2, n^3, n^4) = 0.
\end{array}
\end{equation}
The final equation above is the operator version of the relativistic Klein-Gordon equation in our discrete $(4+4)$-dimension phase space. It is a partial difference equation using the covariant partial difference operator $\Delta^{\#}_{\mu}$.

We shall now propose a partial difference-differential equation for the relativistic Klein-Gordon equation in a discrete phase space continuous time state space of dimension $[(2+2+2)+1]$,
\begin{equation}
\begin{array}{c}
\overrightarrow{\mathbf{\Psi}}  := \phi(n^1, n^2, n^3 ;t) =: \phi(\mathbf{n};t), \\
\\
\mathbf{Q}^{j} \overrightarrow{\mathbf{\Psi}} :=  \delta^{i j} \Delta^{o}_{k} \phi(\mathbf{n};t), \,\,\, \mathbf{Q}^{4} \overrightarrow{\mathbf{\Psi}} :=  t \phi(\mathbf{n};t)\\
\\
\mathbf{P}_{k} \overrightarrow{\mathbf{\Psi}} := -i \Delta^{\#}_{k} \phi(\mathbf{n};t), \,\,\, \mathbf{P}_{4} \overrightarrow{\mathbf{\Psi}} := -i \partial_t \phi(\mathbf{n};t), \\
\\
\left[ \left( \delta^{j k} \mathbf{P}_{j} \mathbf{P}_{k} \right) - \left(\mathbf{P}_{4}\right)^2 +m^2 \mathbf{I} \right] \overrightarrow{\mathbf{\Psi}} =\overrightarrow{\mathbf{0}}, \\
or \\
\left[ \left( \delta^{j k} \Delta^{\#}_{j} \Delta^{\#}_{k} \right) - \partial^2_t- m^2 \right] \phi(\mathbf{n};t) = 0.
\end{array}
\end{equation}

The proof of the exact relativistic invariance of both (20) and (21) is as follows. First, consider the proper orthochronous ten-parameter Poincar\'e group $\mathcal{I}\left[O(3,1)\right]^{+}_{+}$ \cite{WignerII,Weinberg} given by the following equations:
\begin{equation}
\begin{array}{c}
\hat{q}^{\mu} = c^{\mu} +l^{\mu}_{\nu} q^{\nu}, \\
\\
\eta_{\mu \nu} l^{\mu}_{\alpha}l^{\nu}_{\beta}=\eta_{\alpha \beta}, \\
\\
det[l^{\alpha}_{\beta}]=+1,\\
\\
l^4_4 \geq 1.
\end{array}
\end{equation}
Note that the scalar field $\phi(n^1, n^2, n^3, n^4)$ transforms according to the usual tensor analysis \cite{DasTA} as $\phi(\hat{q}^1, \hat{q}^2, \hat{q}^3, \hat{q}^4) = \phi(q^1, q^2, q^3, q^4)$. 

Consider the following unitary operator \cite{DasIV} $U[c^{\mu}, l^{\alpha}_{\beta}]$ which depends upon the ten-parameters of $\mathcal{I}\left[O(3,1)\right]^{+}_{+}$:
\begin{equation}
\begin{array}{c}
U[c^{\mu},l^{\alpha}_{\beta}] := \exp \left\{ -i c^{\mu} \mathbf{P}_{\mu} + \dfrac{i}{4} \omega^{\alpha \beta} \left[  \mathbf{Q}_{\alpha} \mathbf{P}_{\beta}-\mathbf{Q}_{\beta} \mathbf{P}_{\alpha} +\mathbf{P}_{\beta}\mathbf{Q}_{\alpha}-\mathbf{P}_{\alpha}\mathbf{Q}_{\beta}  \right]  \right\}, \\
\omega^{\beta \alpha} =-\omega^{\alpha \beta} .
\\
\end{array}
\end{equation}
In the Schr\"{o}dinger picture, transformations on the state vector $\overrightarrow{\mathbf{\Psi}}$ and operators $\mathbf{Q}^{\mu}$ and $\mathbf{P}^{\nu}$ are induced by the operator $U[c^{\gamma}, l^{\alpha}_{\beta}]$ defined above as follows,
\begin{equation}
\begin{array}{c}
\widehat{\overrightarrow{\mathbf{\Psi}}} = U[c^{\mu}, l^{\alpha}_{\beta}]\overrightarrow{\mathbf{\Psi}}, \\
\widehat{\mathbf{Q}}_{\mu} = \mathbf{Q}_{\mu}, \\
\widehat{\mathbf{P}}_{\nu} = \mathbf{P}_{\nu}.
\end{array}
\end{equation}
However, in the Heisenberg picture, transformations on the state vector $\overrightarrow{\mathbf{\Psi}}$ and operators $\mathbf{Q}^{\mu}$ and $\mathbf{P}^{\nu}$ are induced by the operator $U[c^{\mu}, l^{\alpha}_{\beta}]$ as follows,
\begin{equation}
\begin{array}{c}
\widehat{\overrightarrow{\mathbf{\Psi}}} = \overrightarrow{\mathbf{\Psi}}, \\
\\
\widehat{\mathbf{Q}}_{\mu} = U^{\dagger}[c^{\gamma}, l^{\alpha}_{\beta}] \mathbf{Q}_{\mu}U[c^{\gamma}, l^{\alpha}_{\beta}], \\
\\
\widehat{\mathbf{P}}_{\nu} = U^{\dagger}[c^{\gamma}, l^{\alpha}_{\beta}] \mathbf{P}_{\nu} U[c^{\gamma}, l^{\alpha}_{\beta}].
\end{array}
\end{equation}
Below, we use only the Schr\"{o}dinger picture. The Casimir invariant of $\mathcal{I}\left[O(3,1)\right]^{+}_{+}$ is given by the operator $\eta^{\mu \nu} \mathbf{P}_{\mu}\mathbf{P}_{\nu}$. It commutes with all ten generators $\mathbf{P}_{\mu}$ and $\left(\mathbf{Q}_{\alpha} \mathbf{P}_{\beta}-\mathbf{Q}_{\beta} \mathbf{P}_{\alpha} +\mathbf{P}_{\beta}\mathbf{Q}_{\alpha}-\mathbf{P}_{\alpha}\mathbf{Q}_{\beta}\right)$. Therefore, the Klein-Gordon operator equation in (20) transforms via (24) as
\begin{equation}
\begin{array}{c}
\left[ \left( \eta^{\mu \nu} \widehat{\mathbf{P}}_{\mu} \widehat{\mathbf{P}}_{\nu} \right) +m^2 \mathbf{I} \right] \widehat{\overrightarrow{\mathbf{\Psi}}} =\left[ \left( \eta^{\mu \nu} \mathbf{P}_{\mu} \mathbf{P}_{\nu} \right) +m^2 \mathbf{I} \right] U[c^{\gamma}, l^{\alpha}_{\beta}]\overrightarrow{\mathbf{\Psi}} \\
 \\
=U[c^{\gamma}, l^{\alpha}_{\beta}]\left[ \left( \eta^{\mu \nu} \mathbf{P}_{\mu} \mathbf{P}_{\nu} \right) +m^2 \mathbf{I} \right]\overrightarrow{\mathbf{\Psi}}=\overrightarrow{\mathbf{0}}. \\
\end{array}
\end{equation} 
Therefore, the Klein-Gordon operator equation, the fourth relation in (20), remains invariant under the covariant transformations (24) proving that our Klein-Gordon operator equation is completely special relativistic.

Now we appeal to the partial difference representation of the quantum mechanics provided by last the equation of (20),
\begin{equation}
\left[  \eta^{\mu \nu} \Delta^{\#}_{\mu} \Delta^{\#}_{\nu}  - m^2 \right] \phi(n^1, n^2, n^3, n^4) = 0.
\end{equation}   
The unitary operator $U[c^{\mu}, l^{\alpha}_{\beta}]$ which depends upon the ten-parameters of $\mathcal{I}\left[O(3,1)\right]^{+}_{+}$ may be written as
\begin{equation}
\begin{array}{c}
U[c^{\mu},l^{\alpha}_{\beta}] := \exp \left\{ - c^{\mu} \Delta^{\#}_{\mu} + \dfrac{1}{4} \omega^{\alpha \beta} \left[ \Delta^{o}_{\alpha}  \Delta^{\#}_{\beta}-\Delta^{o}_{\beta} \Delta^{\#}_{\alpha} +\Delta^{\#}_{\beta}\Delta^{o}_{\alpha}-\Delta^{\#}_{\alpha}\Delta^{o}_{\beta}  \right]  \right\}, \\
\\
\hat{\phi}(n^1, n^2, n^3, n^4) :=  \\
\exp \left\{ - c^{\mu} \Delta^{\#}_{\mu} + \dfrac{1}{4} \omega^{\alpha \beta} \left[ \Delta^{o}_{\alpha}  \Delta^{\#}_{\beta}-\Delta^{o}_{\beta} \Delta^{\#}_{\alpha} +\Delta^{\#}_{\beta}\Delta^{o}_{\alpha}-\Delta^{\#}_{\alpha}\Delta^{o}_{\beta}  \right]  \right\} \phi(n^1, n^2, n^3, n^4). 
\end{array}
\end{equation}

In the $[(2+2+2)+1]$-dimensional discrete phase space and continuous time (state space), the state vector 
$\overrightarrow{\mathbf{\Psi}}$ and operators $\mathbf{Q}^{j},\mathbf{Q}^{4},\mathbf{P}^{j},\mathbf{P}^{4}$ are defined in (21) with the partial difference-differential Klein-Gordon equation being given by $\left[  \delta^{j k} \Delta^{\#}_{j} \Delta^{\#}_{k} - \partial^2_t- m^2 \right] \phi(\mathbf{n};t) = 0$. The Klein -Gordon wave function solution transforms under the $\mathcal{I}\left[O(3,1)\right]^{+}_{+}$ group as 
\begin{equation}
\begin{array}{c}
\hat{\phi}(\mathbf{n}; t) := 
\exp \left\{ - c^{j} \Delta^{\#}_{j} - c^{4} \partial_t + \dfrac{1}{4} \left[ \omega^{jk } \Delta^{o}_{j}  \Delta^{\#}_{k}-\Delta^{o}_{k} \Delta^{\#}_{j} +\Delta^{\#}_{k}\Delta^{o}_{j}-\Delta^{\#}_{j}\Delta^{o}_{k}  \right.  \right.
\\
\left.  \left. +\omega^{j4} (t\Delta^{\#}_{j} -\Delta^{o}_{j}\partial_t) \right] \right\} \phi(\mathbf{n}; t). 
\end{array}
\end{equation}
By following similar steps as in (26), one can prove that $\hat{\phi}(\mathbf{n}; t)$ satisfies the  partial difference-differential Klein-Gordon equation  of (21), i.e.
\begin{equation}
\left[  \delta^{j k} \Delta^{\#}_{j} \Delta^{\#}_{k} - \partial^2_t- m^2 \right] \hat{\phi}(\mathbf{n}; t) = 0,
\end{equation}
thereby proving that the partial difference-differential Klein-Gordon equation is exactly invariant under all possible relativistic transformations of the ten parameter Poincar\'e group $\mathcal{I}\left[O(3,1)\right]^{+}_{+}$ .

\section{Concluding Remarks}

In the relativistic quantum field theory, a non-Hermitian scalar field operator $\Phi(n^1, n^2, n^3, n^4)$ or else $\Phi(\mathbf{n};t)$ is a linear operator acting on a Hilbert space vector bundle \cite{DasTA}. The physical principle of the micro-causality \cite{DasV} demands that
$
[\Phi(\mathbf{n};t), \Phi(\hat{\mathbf{n}};\hat{t})]|_{t=\hat{t}}= \mathbf{0} \,\,\, for \,\,\, \mathbf{n} \neq  \hat{\mathbf{n}}
$. However, a similar condition 
$[\Phi(n^1, n^2, n^3, n^4), \Phi(\hat{n}^1, \hat{n}^2, \hat{n}^3, \hat{n}^4)]|_{n^4=\hat{n}^4}= \mathbf{0} $ does not imply the physical principle of micro-causality. Just for this physical consistency, we are justified in working on the more complicated relativistic quantum field theory in the discrete phase space and continuous time arena \cite{DasV}.

We would like to mention the historical fact that Hamilton \cite{Hamilton} introduced the propagation of light wave phenomena using a partial difference equation in lattice space-time as $[\delta^{j k} \Delta_{j} \Delta^{'}_{k}-\Delta_{4} \Delta^{'}_{4}]\phi(n^1, n^2, n^3, n^4) =0$. This equation is of course non-relativistic. In previous work \cite{DasI}, a non-relativistic divergence-free interacting quantum field theory in lattice space-time was similarly introduced.

The quantized Planck oscillators in $[(2+2+2)]$-dimensional phase space create denumerably infinite number of three-dimensional discrete orbits $S^1_{n^1} \times S^1_{n^2} \times S^1_{n^3}$. The Planck oscillators are assumed to be extremely small in size and mass beyond present day experimental detection techniques. However, these oscillators do produce discrete orbits in  $[(2+2+2)]$-dimensional discretized phase space. The second quantized relativistic Klein-Gordon operator wave field $\Phi(\mathbf{n};t)$ representing other physical particles like mesons, that are distinct from the first quantized oscillators, can temporarily, or for all time, occupy any of the geometrical objects  $S^1_{n^1} \times S^1_{n^2} \times S^1_{n^3}$. Bosonic particles can cohabit with a first quantized Planck oscillator in any one of the orbits $S^1_{n^1} \times S^1_{n^2} \times S^1_{n^3}$. Furthermore, some of the configurations $S^1_{n^1} \times S^1_{n^2} \times S^1_{n^3}$ can remain totally empty of Bosons for a period of time. In the $S^{\#}$-matrix theory \cite{DasVI} of interacting Fermion and Boson fields, it was assumed that a field quanta can jump from one configuration $S^1_{n^1} \times S^1_{n^2} \times S^1_{n^3}$ into another one completely ignoring the Planck oscillators.

In future work, we intend to improve the physical side of this theory by changing the three dimensional geometrical objects $S^1_{n^1} \times S^1_{n^2} \times S^1_{n^3}$ into six-dimensional objects $\bar{S}_{n^1} \times \bar{S}_{n^2} \times \bar{S}_{n^3}$ inside the $(2+2+2)$-dimensional discrete phase space. Here, $\bar{S}^1_{n^j}$ denotes a Peano circle \cite{Clark} of dimension two inside the $(2+2+2)$-dimensional discrete phase space. We can identify $\bar{S}_{n^1} \times \bar{S}_{n^2} \times \bar{S}_{n^3}$  as a six-dimensional phase cell inside $\mathbb{R}^2 \times \mathbb{R}^2 \times \mathbb{R}^2$. We shall elaborate on this topic exhaustively in the succeeding part II of this paper entitled \textit{Discrete phase space and continuous time relativistic quantum mechanics II: Peano circles and hyper-tori like phase cells}.

\end{document}